\documentclass[preprint,aps]{revtex4}
\usepackage{graphicx}

\begin{document}

\title{Rare $B_{s}\rightarrow \nu \overline{\nu }\gamma$ decay
beyond the standard model}
\author{O. Cakir}
\affiliation{Ankara University, Faculty of Sciences, Department of Physics, \\
  06100, Tandogan, Ankara, Turkey.}
\author{B. Sirvanli}
\affiliation{Gazi University, Faculty of Arts and Sciences, Department of Physics, \\
  06500, Teknikokullar, Ankara, Turkey.}

\begin{abstract}
Using the most general model independent form of the effective
Hamiltonian the rare decay $B_{s}\rightarrow\nu\bar{\nu}\gamma$
is studied. The sensitivity of the
photon energy distribution and branching ratio to new Wilson
coefficients are investigated.
\end{abstract}


\maketitle

\section{Introduction}

The experimental observation of the $b\rightarrow s\gamma $ [1]
and $B\rightarrow X_s\gamma $ [2] processes opened a window to
study possible rare $B$ meson decays, which will give important
information about Cabibbo-Kobayashi-Maskawa (CKM) matrix elements.
In the Standard Model rare $B_{s}\rightarrow \nu \overline{\nu }$
decay is forbidden due to the helicity conservation. When photon
is emitted from structure dependent part (see below) the helicity
conservation allows the decay $B_{s}\rightarrow \nu \overline{\nu
}\gamma $. Therefore, the investigation of the $B_{s}\rightarrow
\nu \overline{\nu }\gamma $ rare decay becomes interesting.

The main interest in studying $B$ meson decays is to test the
Standard Model (SM) predictions at loop level and extract new
physics effects beyond the SM. The loop induced rare decays
$B_{s}(B_{d})\rightarrow \nu \overline{\nu } \gamma $ \ within the
SM have been studied using constituent quark model and pole model
and the branching ratios are found to be $10^{-8}$ for $
B_{s}\rightarrow \nu \overline{\nu }\gamma $\ and $10^{-9}$ $
B_{d}\rightarrow \nu \overline{\nu }\gamma $ [3].

The decay rate of the $B_s\rightarrow \nu \overline{\nu }\gamma $
might have an enhancement comparing with the pure leptonic modes
of $B$ meson. The new physics effects beyond SM can be also probed
by studying this $B_{s}\rightarrow \nu \overline{ \nu }\gamma $
rare decay properties.

In this paper, we study the sensitivity of the physically
measurable quantities, branching ratio, photon energy
distribution, to the new physics effects using the most general
form of the effective Hamiltonian. The effects of the new Wilson
coefficients $C_X$ on branching ratio and photon energy
distribution is also investigated.

\section{Effective Hamiltonian}

The most general model independent form of the effective
Hamiltonian for the process $b\rightarrow q\nu \overline{\nu }$
can be written in the following form [4,5];
\begin{eqnarray}
H_{eff}&=&\frac{G_F\alpha V_{tb}V_{ts}^{*}}{4\sqrt{2}\sin ^2\theta
_w} \{C_{LL}^{tot}\,\overline{s}\,\gamma _\mu \,(1-\gamma
_5)\,b\,\overline{\nu } \,\gamma ^\mu \,(1-\gamma _5)\,\nu
+C_{LR}^{tot}\,\overline{s}\,\gamma _\mu \,(1-\gamma
_5)\,b\,\overline{\nu }\,\gamma ^\mu \,(1+\gamma _5)\,\nu \nonumber\\
&+&C_{RL}\,\overline{s}\,\gamma _\mu \,(1+\gamma
_5)\,b\,\overline{\nu }\,\gamma ^\mu \,(1-\gamma _5)\,\nu
+C_{RR}^{}\, \overline{s}\,\gamma _\mu \,(1+\gamma
_5)\,b\,\overline{\nu
}\,\gamma ^\mu \,(1+\gamma _5)\,\nu \nonumber \\
&+&C_{LRLR}^{}\,\overline{s}\,(1+\gamma _5)\,b\, \overline{\nu
}\,\,(1+\gamma _5)\,\nu +C_{RLLR}^{}\,\overline{s}\,(1-\gamma
_5)\,b\,\overline{\nu }\,\,(1+\gamma _5)\,\nu \nonumber \\
&+&C_{LRRL}^{}\,\overline{s}\,(1+\gamma _5)\,b\, \overline{\nu
}\,\,(1-\gamma _5)\,\nu +C_{RLRL}^{}\,\overline{s}\,(1-\gamma
_5)\,b\,\overline{\nu }\,\,(1-\gamma _5)\,\nu \nonumber \\
&+&C_T\,\overline{s}\,\sigma _{\mu \nu }\,b\,%
\overline{\nu }\,\,\sigma ^{\mu \nu }\,\nu +i\,C_{TE}\,\in ^{\mu
\nu \alpha \beta }\overline{s}\,\sigma _{\mu \nu
}\,b\,\overline{\nu }\,\sigma _{\alpha \beta }\,\nu \}
\end{eqnarray}
where the projection operators $L$ and $R$ in Eq.(1) are defined
as $L=(1-\gamma _5)/2$, $R=(1+\gamma _5)/2$, and $C_X$ are the
coefficients of the four-Fermi interactions. In this work, we
restrict ourselves by considering only Dirac neutrino in order to
avoid additional lepton-number-violating operators. Furthermore,
the states $\nu_L$ and $\nu_R$ are well separated in the massless
Dirac neutrino case. However, for the massive neutrino case we can
change chirality with help of the mass, and therefore in this case
it is necessary the right-handed neutrino to be heavy. In general,
the Wilson coeeficients for $b\rightarrow sl^{+}l^{-}$ and
$b\rightarrow s\nu\bar{\nu}$ processes are different. However,
operator structures in both case must be the same, since for
massive neutrino case charged lepton and neutrino except their
electric charge are the same with respect to the SU(2). First four
terms in the effective Hamiltonian are the vector interactions.
The interaction terms with coefficients $C_{LL}^{tot}$ and
$C_{LR}^{tot}$ which are present in the SM are given in the form,
\begin{eqnarray*}
C_{LL}^{SM} &=&C_9^{eff}-C_{10}\\
C_{LR}^{SM}&=&C_9^{eff}+C_{10}.
\end{eqnarray*}
The contributions from the new physics can be described by
redefining the Wilson coefficients as
$C_{LL(LR)}^{tot}=C_{LL(LR)}^{SM}-C_X$. The coefficients
$C_{LRLR},C_{RLLR},C_{LRRL}$ and $C_{RLRL}\;$ describe the
scalar type interactions which disappear in our calculations for $%
B_s\rightarrow \nu \overline{\nu }\gamma $ process. The remaining
last two coefficients in Eq.(1), correspond to tensor type
interactions. In general, these operators are possible and they
can appear from the exchange of spin-2 particles. It is obvious
that the effects of the $C_T$ and $C_{TE}$ operators will be
larger, because there are more Lorentz indices, which means that
summing over them will lead to larger effects. In order to
calculate the matrix element for $B_s\rightarrow \nu \overline{\nu
}\gamma $ decay we use the general form of the effective
Hamiltonian and the standard definitions for the matrix elements
[6,7,8] ;
\begin{eqnarray}
\langle \;\gamma \,(q)\left| \overline{s}\,\gamma _\mu \,(1\mp
\gamma _5)\,b\right| B(p_B)\,\,\rangle =\frac e{m_B^2}\{\in _{\mu
\nu \lambda \sigma }\varepsilon ^{*\nu }p^\lambda q^\sigma
g\,(p^2)\pm \,i\,\left[ \varepsilon ^{*\mu }(pq)-(\varepsilon
^{*}p)q^\mu \right] \,f(p^2)\}
\end{eqnarray}
\begin{equation}
 \langle \;\gamma \,(q)\left| \overline{s}\,\,\sigma _{\mu
\nu }\,b\right| B(p_B)\,\,\rangle =\frac e{m_B^2}\in _{\mu \nu
\lambda \sigma }\left[ G\,\varepsilon ^{*\lambda }q^\sigma
+H\,\varepsilon ^{*\lambda }p^\sigma +N\,(\varepsilon
^{*}p)\,p^\lambda \,q^\sigma \right]
\end{equation}
\begin{equation}
 \langle
\;\gamma \,(q)\left| \overline{s}\,\,(1\mp \gamma _5)\,b\right|
B(p_B)\,\,\rangle =0
\end{equation}
where $\varepsilon _\mu ^{*}$ is the polarization vector of the photon. The $%
p$,$q$ and $p_B$ are the transfer momentum, photon momentum and
the momentum of $B$ meson, respectively. Using Eqs. (2), (3) and
(4) the matrix element for the process can be calculated as
follows:
\begin{eqnarray}
M&=&\frac{\alpha \,G_F}{4\sqrt{2}}V_{tb}V_{ts}^{*}\frac
e{m_B^2}\{\overline{ \,\nu }\,\gamma ^\mu \,(1-\gamma _5)\,\nu
\,\left[ A_1\in _{\mu \nu \alpha \beta }\varepsilon ^{*\nu
}p^\alpha q^\beta +i\,A_2(\varepsilon _\mu ^{*}\,(pq)-(\varepsilon
^{*}p)q_\mu
^{})\right]\nonumber \\
&&\quad +\overline{\,\nu }\,\gamma ^\mu \,(1+\gamma _5)\,\nu
\,\left[ B_1\in _{\mu \nu \alpha \beta }\varepsilon ^{*\nu
}p^\alpha q^\beta +i\,B_2(\varepsilon _\mu ^{*}\,(pq)-(\varepsilon
^{*}p)q_\mu ^{})\right] \nonumber \\
&+&\,i\,\in _{\mu \nu \alpha \beta } \overline{\nu }\,\,\sigma
^{\mu \nu }\,\nu \,\,\,\left[ G\,\varepsilon ^{*\alpha }q^\beta
+H\,\varepsilon ^{*\alpha }p^\beta +N\,(\varepsilon
^{*}p)\,p^\alpha \,q^\beta \right]\nonumber \\
&+&\,i\,\overline{\nu }\,\,\sigma _{\mu \nu }\,\nu \,\left[
G_1(\varepsilon ^{*\mu }q^\nu -\varepsilon ^{*\nu }q^\mu
)+H_1(\varepsilon ^{*\mu }p^\nu -\varepsilon ^{*\nu }p^\mu
)+N_1(\varepsilon ^{*}p)(p^\mu q^\nu -p^\nu q^\mu )\right] \}
\end{eqnarray}
where we have used the definitions [6];
\begin{eqnarray}
A_1^{} &=&(C_{LL}^{tot}+C_{RL})\,\,g,\qquad \qquad \qquad \qquad
A_2^{}=(C_{LL}^{tot}-C_{RL})\,\,f, \nonumber \\
B_1^{} &=&(C_{LR}^{tot}+C_{RR})\,\,g,\qquad \qquad \qquad \qquad
B_2^{}=(C_{LR}^{tot}-C_{RR})\,\,f. \nonumber \\
G &=&4C_T\,g_1,\qquad \qquad \qquad
N=-4\,C_T\,\frac{(f_1+g_1)}{p^2},\qquad
\qquad \qquad H=N\,(pq), \nonumber \\
G_1 &=&-8C_{TE}\,g_1,\qquad \quad \qquad
N_1=8\,C_{TE}\,\frac{(f_1+g_1)}{p^2} ,\quad \quad \quad
H_1=N_1\,(pq).
\end{eqnarray}
with $p^2=m_B^2(1-x),\,pq=m_B^2x/2,\,x=2E_\gamma/m_B$ where
$E_\gamma $ is the photon energy. The double differential decay
width of the $B_s\rightarrow \nu \overline{\nu }\gamma $ process
in the rest frame of the $B$ meson can be written in the form
\begin{eqnarray}
\frac{d\,\Gamma }{d\,x\,d\,E_1} &=&\frac 1{128\,\pi ^3}\left|
M\right| ^2
\end{eqnarray}
The bounds of the final neutrino energy $E_1$ and the
dimensionless parameter $x$ for photon energy are determined from
the following inequalities
\begin{eqnarray}
\frac{m_B}2-E_\gamma &\leq &E_1\leq \frac{m_B}2,\qquad \qquad
\qquad 0\leq x\leq 1
\end{eqnarray}
In our calculations, we consider hard photon in the process
$B_s\rightarrow \nu \overline{\nu }\gamma $ . For the experimental
observability of photon we impose a cut on the minimum energy to
be greater than 25 MeV which corresponds to $x\geq 0.01$ [6,8,9].
We integrate the differential decay width over the neutrino energy
$E_1$ to get the photon energy distribution
\begin{eqnarray}
\frac{d\,\Gamma }{d\,x\,}&=&-\,\left| \frac{\alpha
\,G_F}{4\sqrt{2}} V_{tb}V_{ts}^{*}\right| ^2\,\frac \alpha {(2\pi
)^3}\,\frac \pi 4\,m_B\,x^3\,\{\,-4\,\left[ \left| H_1\right|
^2(1-x)+Re(G_1H_1^{*})\,x\right] \,\frac{1-x}{x^2} \nonumber \\
&-&4\left[ \left| H\right| ^2(1-x)+Re(GH^{*})\,x\right] \frac{
1-x}{x^2}+\frac 13\,m_B^2\,\left[
2Re(GN^{*})+m_B^2\left| N\right| ^2(1-x)\right] \,(1-x) \nonumber \\
&+&\frac 13\,m_B^2\,\left[ 2Re(G_1N_1^{*})+m_B^2\left|
N_1\right| ^2(1-x)\right] \,(1-x) \nonumber \\
&-&\frac 23m_B^2\left[ (\left| A_1\right| ^2+\left| A_2\right|
^2+\left| B_1\right| ^2+\left| B_2\right| ^2)\,(1-x)\right]
\,-\frac 43\,(\left| G\right| ^2+\left| G_1\right| ^2)\}
\end{eqnarray}
The form factors $f,g,f_1$ and $g_1$ ,which we have used in our
numerical calculations, appearing in Eq.(6) can be obtained in the
framework of light cone QCD sum rules given in [7,8] and their $x$
dependence with a good accuracy:
\begin{eqnarray*}
f(x) &=&\frac{0.8\,GeV}{(1-\frac{(4.8)^2(1-x)}{(6.5)^2})^2},\qquad
\qquad
\qquad \qquad g(x)=\frac{1\,GeV}{(1-\frac{(4.8)^2(1-x)}{(5.6)^2})^2}, \\
&& \\
f_1(x) &=&\frac{0.68\,GeV^2}{(1-\frac{(4.8)^2(1-x)}{30})^2},\qquad
\qquad \qquad \qquad
g_1(x)=\frac{3.74\,GeV^2}{(1-\frac{(4.8)^2(1-x)}{40.5})^2}
.\qquad \qquad (10) \\
&&
\end{eqnarray*}

\section{Results and Discussions}

We use the main input parameters $m_B=5.28GeV,\,\left|
V_{tb}V_{ts}^{*}\right| =0.045,\alpha ^{-1}=137,G_F=1.17\times
10^{-5}GeV^{-2}$ in the numerical results. For the Wilson
coefficients we take the values $C_9^{eff}=4.344$ and
$C_{10}=4.6242$ given in the [6,10].

\begin{figure}
\includegraphics{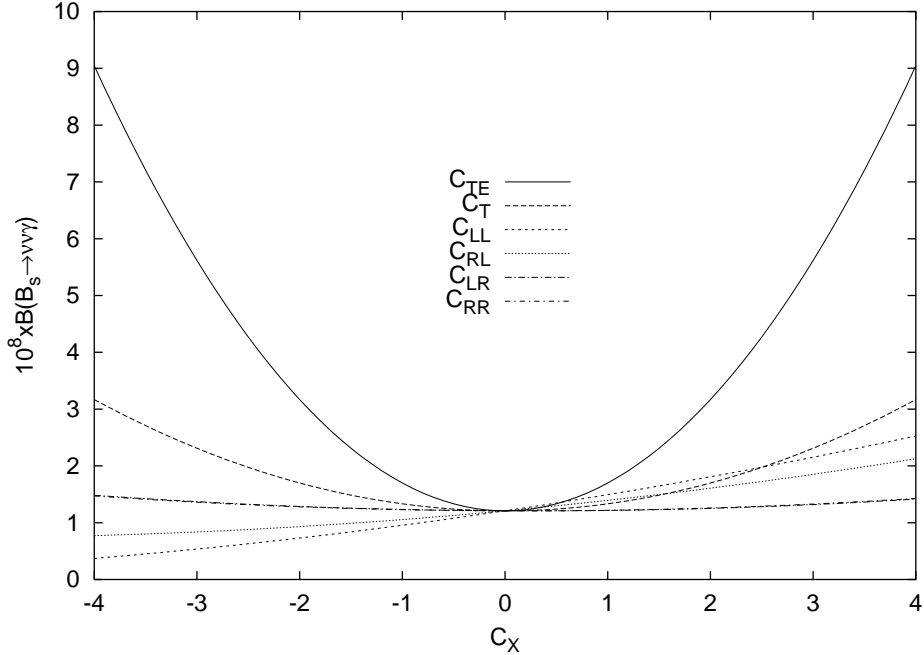} \caption{The branching ratio of the rare
$B_s\rightarrow \nu \overline{\nu } \gamma $ decay depending on
the new Wilson coefficients $C_X$ for the parameter cut $x_{\min
}=0.01$.} \label{fig1}
\end{figure}

In this work, we assume that all new Wilson coefficients $C_X$ are
real and vary in the region $-4\leq $ $C_X\leq +4$. The
experimental bounds on the branching ratios of the $B\rightarrow
K^{\star}\mu^{+}\mu^{-}$ and $B_{s}\rightarrow \mu^{+}\mu^{-}$
[11] suggest that this is the right order of magnitude range for
the vector and scalar interactions coefficients. Therefore, we
assume that all new Wilson coefficients change in this range. The
integrated branching ratio for the rare $B_{s}\rightarrow \nu
\overline{\nu }\gamma $ decay depending on the new Wilson
coefficients $C_T,C_{TE},C_{RR},C_{RL},C_{LL},C_{LR}$ is plotted
in Fig. \ref{fig1}. It is clear from this figure that branching
ratio increases when all new Wilson coefficients $C_{X}>0$
increase, for
example, in the region $-4\leq $ $C_T,$ $%
C_{TE}\leq 0$ branching ratio decreases and it increases in the region $%
0\leq $ $C_T$ $,C_{TE}\leq +4$. Furthermore, the branching ratio
remain approximately unchanged when the coefficients
$C_{RR},C_{LR}$ varies and there is no sensitivity to these
coefficients. The branching ratio weakly depends on the
$C_{RR},C_{RL},C_{LR}$ coefficients.

\begin{figure}
\includegraphics{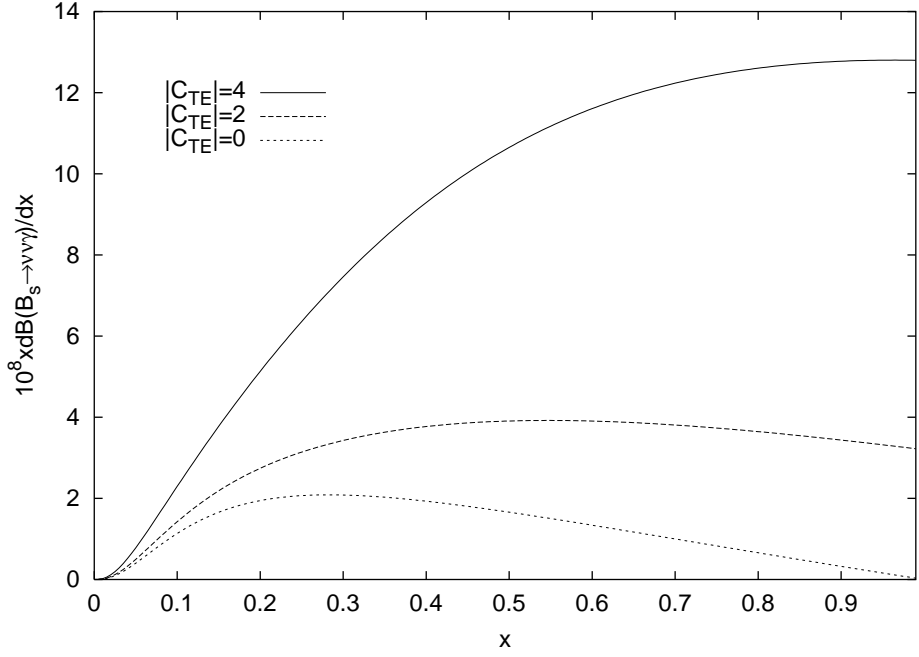} \caption{The differential branching ratio for
rare $B_s\rightarrow \nu \overline{\nu }\gamma $ depending on the
dimensionless variable $x=2E_\gamma /m_B$ for the different values
of tensor interaction coefficient $C_{TE}$ } \label{fig2}
\end{figure}

However, one can conclude that the branching ratio is more
sensitive to the tensor type $C_T$ and $C_{TE}$ coefficients. We
find the branching ratio $ B(B_s\rightarrow \nu \overline{\nu
}\gamma )=1.2\times 10^{-8}$ for new Wilson coefficients are set
to zero. The branching ratio for the contact interactions give
symmetrical distribution with respect to zero. Measuring this
branching ratio can give an information about the magnitude of
this type tensor interactions, see Fig. \ref{fig2}. In addition,
the $C_{RL}$ and $C_{LL}$ distribution can give an opportunity to
detect the sign of these coefficients. We see that in dependence
of the sign of the tensor interaction $C_{TE}$ differential
branching ratio can be larger or smaller than SM prediction .

\begin{figure}
\includegraphics{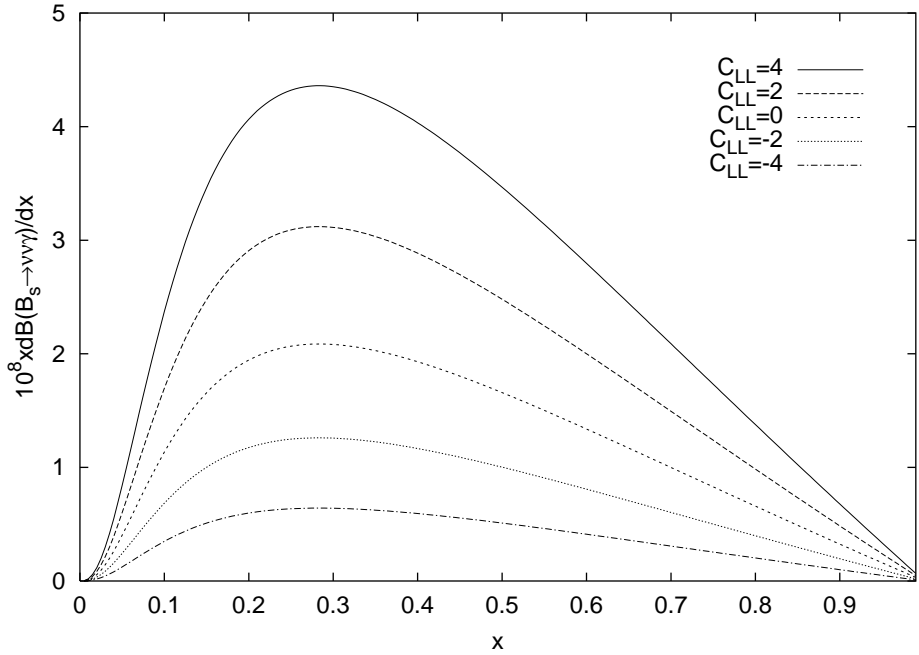} \caption{The differential branching ratio for
rare $B_s\rightarrow \nu \overline{\nu }\gamma $ depending on the
dimensionless variable $x=2E_\gamma /m_B$ for the different values
of vector interaction coefficient $C_{LL}$ } \label{fig3}
\end{figure}

The photon energy distribution can also give information about new
physics effects. In Fig. \ref{fig3}, we present the differential
branching ratio for the rare $B_{s}\rightarrow \nu \overline{\nu
}\gamma $ decay as a function of dimensionless variable $x$ for
different values of coefficient $C_{LL}$. In this view the
differential branching ratio measurement could give important
information about the sign of new Wilson coefficient. The higher
sensitivity can be obtained when the photon energy reaches ~$\sim
0.6GeV$.

In conclusion, using a general model independent effective
Hamiltonian for the process $B_s\rightarrow \nu \overline{\nu
}\gamma $ , the branching ratio and the photon energy distribution
are found to be sensitive to the existence of new physics beyond
the SM. Within a reasonable range of branching ratios, it would be
possible to detect the rare processes in the future B-factories.
At planning LHC-B and B TeV hadronic machines $10^{11}-10^{12}$
$bb$ pair per years [12] will be produced. Therefore, the number
of expected events are $N\approx 10^3-10^4$, which quite
detectable this decay in the above mentioned colliders. Note that
signature of this decay will be single photon and missing energy.

\acknowledgments

We would like to thank T.M. Aliev and S. Sultansoy for useful
discussions. This research is partially supported by Turkish
Planning Organization (DPT) under the Grant No 2002K120250.

\end{document}